\documentstyle[twocolumn,prb,floats,aps,psfig]{revtex}
\newcommand{\mywidth}{3.8in}

\begin{document}
\draft

\title{Finite--size scaling of the helicity modulus of the two--dimensional
O(3) model}
\author{N. Schultka}
\address{Institut f\"ur Theoretische Physik, Technische
Hochschule Aachen, D--52056 Aachen, Germany}
\date{\today}
\maketitle
\begin{abstract}
Using Monte Carlo methods, we compute the finite--size scaling function of 
the helicity modulus $\Upsilon$ of the 
two--dimensional O(3) model  and 
compare it to the low temperature expansion
prediction. From this, we estimate the range of validity for the leading 
terms of the 
low temperature expansion of the finite--size scaling function and for the
low temperature expansion of the correlation length.
Our results strongly suggest that a Kosterlitz--Thouless
transition at a temperature $T > 0$ is extremely unlikely in this model.
\end{abstract}
\pacs{ }
We report a computation of the finite--size scaling
function of the helicity modulus of the two--dimensional O(3) model on the
lattice in the presence of periodic boundary conditions. 
The
lattice hamiltonian of the O(3) model is defined as follows:
\begin{equation}
{\cal H}=-J \sum_{\langle i,j \rangle} \vec{s}_i \cdot \vec{s}_j, \label{ham}
\end{equation}
where the sum runs over nearest neighbors, the pseudospin 
$\vec{s}=(s_x,s_y,s_z)$ with $s_x^2+s_y^2+s_z^2=1$, and $J>0$ 
sets the energy scale.
The continuum model and its lattice version (\ref{ham}) are supposed to 
describe the same long wave length physics,
i.e. they exhibit asymptotic freedom or in other words, their critical
temperature is zero\cite{POLYAKOV,LOWT1,LOWT}. This is in accord with the
Mermin--Wagner theorem\cite{MERMIN} which forbids a transition to a state of
long range order in a two--dimensional system of O(N) symmetry with N$>1$ 
at a temperature $T>0$. However, a Kosterlitz--Thouless like
transition is still allowed as is well known for the 
two--dimensional O(2) model\cite{THOULESS}. Indeed, there have been
arguments in favor of a Kosterlitz--Thouless transition in the 
two--dimensional O(3) model\cite{SEILER} which
have recently been subject to various Monte Carlo 
investigations\cite{SOKAL,ALLES} focusing on the low temperature behavior of
the correlation length.
An easily accessible quantity signalling the Kosterlitz--Thouless transition
is the helicity modulus. If a Kosterlitz--Thouless transition occurs in the
model, on an infinite lattice the helicity modulus 
exhibits a jump at the transition temperature
$T_c>0$ from zero to a finite value and remains nonzero for temperatures 
$T< T_c$\cite{NELSON}. However,
our result for the finite--size scaling function 
suggests that the helicity modulus
vanishes at all temperatures $T>0$ on infinite lattices, thus a 
Kosterlitz--Thouless transition is very unlikely to occur in the 
two--dimensional O(3) model.

We compare our result for the finite--size scaling function to the
low temperature expansion prediction of Br\'ezin et al.\cite{BREZIN} obtained
for the O(3) nonlinear $\sigma$ model and given by
\begin{equation}
\frac{\Upsilon(T,L)}{T}=\frac{1}{2\pi}\left(\ln \frac{\xi_{\Upsilon}(T)}{L}+
     \ln \ln \frac{\xi_{\Upsilon}(T)}{L} \right) +O(1). \label{scale}
\end{equation}
Here $\xi_{\Upsilon}$ denotes a correlation length of the system which 
behaves as\cite{LOWT}
\begin{equation}
\xi_{\Upsilon}(T) =C_{\Upsilon} (2\pi/T)^{-1}\, \exp(2\pi/T). \label{xilowt}
\end{equation}
The prefactor $C_{\Upsilon}$ cannot be determined within perturbation theory.
Note that the scaling form (\ref{scale}) is only valid in the limit
$\xi_{\Upsilon}(T) \gg L$. 

Numerically we determine the scaling function 
as follows. We compute the helicity modulus at various temperatures and for
different lattice sizes and plot $\Upsilon(T,L)/T$ versus 
$\ln (L C_{\Upsilon}/\xi_{\Upsilon}(T))$. In the range where 
the low temperature expansion predictions Eq.(\ref{scale}) and
Eq.(\ref{xilowt}) hold, we expect our scaled data to collapse onto a single
curve.
Thus, we are able to estimate the range in the variable 
$z=\ln (L C_{\Upsilon}/\xi_{\Upsilon}(T))$ where the leading 
behavior (\ref{scale}) sets in and the temperature range where
the low temperature expression (\ref{xilowt}) is valid.
For the latter let us assume that the low 
temperature expansion (\ref{xilowt}) were not valid above a certain
temperature $T_{+}$ say. Then we would
see deviations from the scaling curve for values of 
$z$ larger than $\ln (L C_{\Upsilon}/\xi_{\Upsilon}(T_{+}))$.

For our Monte Carlo simulations, we use the hamiltonian
(\ref{ham}) to compute the helicity modulus of the O(3) model on 
$L \times L$ lattices with $L=10,40,60,100,200$ employing Wolff's 
1--cluster algorithm\cite{WOLFFCL}. In order to avoid
boundary effects we apply periodic boundary conditions. Though the
leading terms of the scaling expression (\ref{scale}) were derived for 
fixed spin boundary conditions, the scaling form (\ref{scale})
remains valid for periodic boundary
conditions as well. The definition of the helicity modulus along the 
space direction $\mu$ in the presence
of periodic boundary conditions 
is (for the derivation
follow the steps outlined in Refs.\cite{CHAKRAVARTY})
\begin{eqnarray}
\Upsilon_{\mu}(T,L) & = &
\frac{1}{L^2} \left\langle \sum_{i} 
\vec{s}_i \cdot \vec{s}_{i+\mu} \right. \nonumber \\
& - & \frac{1}{2T} \sum_{[l,m]} \left. \left( 
\sum_{i} s_i^l \wedge s_{i+\mu}^m \right)^2
\right\rangle, \label{helmod} \\
s_i^l \wedge s_{i+\mu}^m & = & s_i^l s_{i+\mu}^m - s_i^m s_{i+\mu}^l, \nonumber
\end{eqnarray}
where $s^l$ and $s^m$ denote two different components of the spin $\vec{s}$.
The summation is over all possible pairs of spin 
components, denoted by $\sum_{[l,m]}$, and over all lattice sites, denoted
by $\sum_{i}$. The symbol $i+\mu$ means the adjacent lattice site of $i$ in the
space direction $\mu$.
In the following we will write $\Upsilon$ instead of $\Upsilon_{\mu}$ as our
lattice is isotropic. Our definition of the helicity modulus (\ref{helmod})
ensures $\lim_{T\rightarrow 0} \Upsilon(T,L)=1$ which agrees with 
Eq.(\ref{scale}).
\begin{figure}[htp]
 \centerline{\psfig {figure=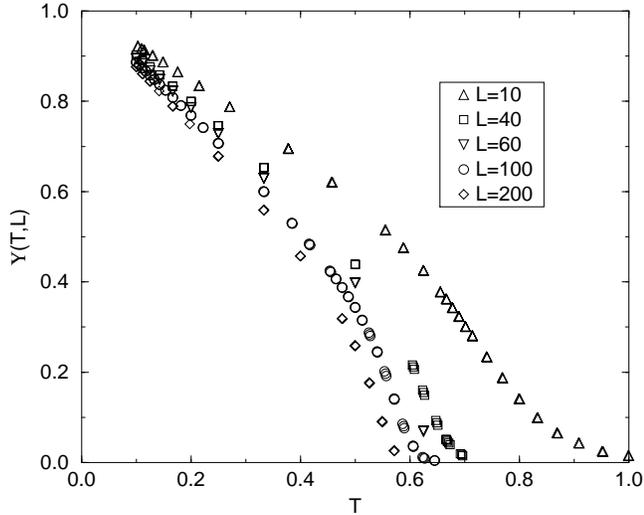,width=\mywidth}}
 \caption{\label{fig1} The helicity modulus $\Upsilon(T,L)$ as a function of 
  temperature $T$ for various lattices $L \times L$. The error bars are smaller
  than the symbols.}
\end{figure}
\begin{figure}[htp]
 \centerline{\psfig {figure=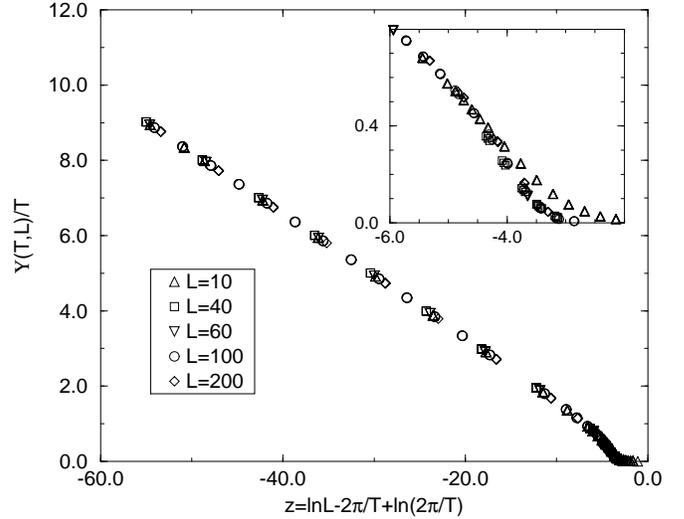,width=\mywidth}}
 \caption{\label{fig2} Scaling plot $\Upsilon(T,L)/T$ versus
 $z=\ln L-2\pi/T$ $+\ln(2\pi/T)$.
 The error bars are smaller than the symbols.}
\end{figure}

Fig.\ref{fig1} shows our data of the helicity modulus as a function of 
temperature for various lattice sizes. From this data it is impossible to
draw conclusions about the behavior of the helicity modulus of the O(3) model
on an infinite lattice. For this reason we have examined
the finite--size scaling function in the manner described above.

In Fig.\ref{fig2} we plot $\Upsilon(T,L)/T$ versus
$z=\ln L -2\pi/T+\ln(2\pi/T)$ using the data of Fig.\ref{fig1}. 
The data collapse onto a single curve with very little scatter
for values $z<-5$, only for $z>-5$ do the data for 
$\Upsilon(T,10)/T$ begin to deviate from the single scaling curve 
(cf. the inset of 
Fig.\ref{fig2}), signalling the break down 
of the expression (\ref{xilowt}). 
The corresponding temperature is $T \approx 0.657$, i.e.
for temperatures $T$ smaller than $0.657$ the low temperature expansion result
(\ref{xilowt}) should be a good description of the temperature dependence
of the correlation length. 
This temperature is larger than the value $0.37$, obtained 
in Refs.\cite{SOKAL,HASENFRATZ,WOLFF,APOSTOLAKIS} where the correlation
length was computed directly via its second moment definition, and 
obtained by different methods in Refs.\cite{NEUHAUS}.

So far we have only shown that our data of the helicity modulus computed
at various temperatures and for different lattice sizes obey scaling. 
Now we wish to
compare the scaled data in Fig.\ref{fig2} with the low temperature
prediction Eq.(\ref{scale}). To this end
we fit the expression
\begin{equation}
f(z)=a_1 (a_2-z+\ln(a_2-z)), \label{fit}
\end{equation}
to our scaled data treating $a_1$ and $a_2=\ln C_{\Upsilon}$ as fit parameters.
In principle we should add a constant
$a_3$ to $f(z)$ to allow for the $O(1)$ term in Eq.(\ref{scale}),
however $a_3$ turned out to be zero within error bars, therefore we set $a_3=0$
for our fits.
\begin{figure}[htp]
 \centerline{\psfig {figure=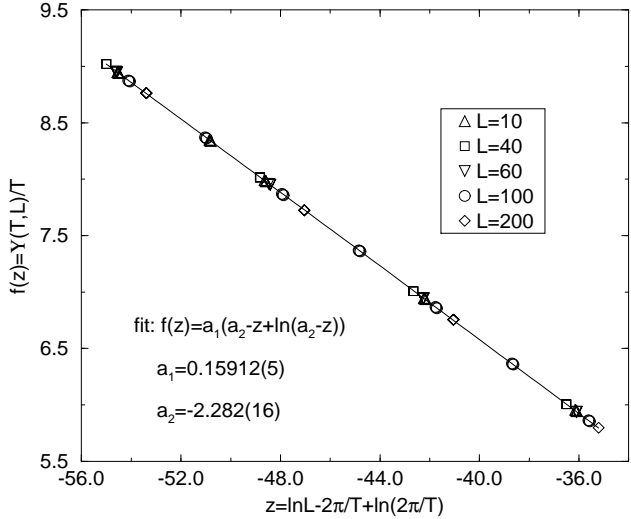,width=\mywidth}}
 \caption{\label{fig3} Fit of the expression (\protect\ref{fit}) to the scaled
 data of Fig.\protect\ref{fig2} in the range $-56 < z < -35$ (70 
 data points).
 The error bars are smaller than the symbols.}
\end{figure}
\begin{table}[htp] \centering
 \begin{tabular}{|l|l|l|l|l|}
 \multicolumn{1}{|c|}{data points}  & \multicolumn{1}{c|}{$a_1$} &
\multicolumn{1}{c|}{$a_2$} & \multicolumn{1}{c|}{$\chi^2$ p.d.f.} &
\multicolumn{1}{c|}{$Q$} \\
\hline 
20 & 0.1590(4) & -2.23(14) & 0.50 & 0.96 \\
40 & 0.1592(2) & -2.32(5)  & 0.64 & 0.96 \\
60 & 0.15917(7)& -2.30(2)  & 0.93 & 0.63 \\
70 & 0.15912(5)& -2.28(2)  & 1.11 & 0.25 \\
80 & 0.15905(4)& -2.26(1)  & 1.09 & 0.27 \\
100& 0.15888(3)& -2.213(6) & 1.38 & 0.0076 \\
 \end{tabular}
\caption{\label{ta1} Fitting results of expression (\protect\ref{fit})
to the scaled data shown in Fig.\protect\ref{fig2}. $\chi^2$ is per degree
of freedom and $Q$ is the quality of the fit.}
\end{table}

Table \ref{ta1} contains our fitting results. We varied the number of
data points in the fit to check the stability of our fitting results and to
estimate the range of validity of the low
temperature result (\ref{scale}) in terms of the variable $z$.
We estimate this range from the quality of fit $Q$ 
(see e.g. Eq.(6.2.18) of Ref.\cite{RECIPE});
the closer $Q$ is to 1, the more reliable are the 
values obtained for the fit parameters $a_1$ and $a_2$.
The value of the 
parameter $a_1$ agrees rather well with the predicted value of 
$1/(2\pi)=0.15915...$,
compare the results of Ref.\cite{CAFFAREL} where the best value 
$a_1=0.162(4)$ was obtained. Fig.\ref{fig3} shows the fitting results of the
fit of the form (\ref{fit}) to our scaled data in the range $-56 < z < -35$,
i.e. including 70 data points. 

Since the quality of the fit decreases by an order of magnitude when about
100 data points are included in the fit (cf. Table \ref{ta1}), we are on the 
safe side assuming that
formula (\ref{scale}) is valid for values of $z$ in the interval 
$(-\infty,-35]$ (cf. Fig.\ref{fig3}), 
i.e. expression (\ref{scale}) holds for temperatures and
lattice sizes that fulfill
\begin{equation}
\ln \left( \frac{T}{2\pi L}\exp\frac{2\pi}{T} \right) > 35. \label{range}
\end{equation}

Our results suggest that the correlation length entering the finite--size
scaling form (\ref{scale}) depends on the temperature according to 
Eq.(\ref{xilowt})
with $C_{\Upsilon}=0.102$.
The prefactor $C_{\Upsilon}$ is about eight times larger than the 
prefactor $0.0125$\cite{SOKAL,HASEN}
obtained from
the second moment definition of the correlation length $\xi$\cite{SOKAL,HASEN}.
Such a comparatively large prefactor was also found in Ref.\cite{BILLOIRE}. 
The difference in the prefactors
is no surprise to us since we can consider the scaling expression 
(\ref{scale}) as providing an alternative definition of the correlation length 
with a prefactor different from that of the second moment definition. 
\begin{table}[htp] \centering
 \begin{tabular}{|l|l|l|l|l|}
 \multicolumn{1}{|c|}{data points}  & \multicolumn{1}{c|}{$a_1$} &
\multicolumn{1}{c|}{$a_2$} & \multicolumn{1}{c|}{$\chi^2$ p.d.f.} &
\multicolumn{1}{c|}{$Q$} \\
\hline 
20 & 0.1622(4) & 0.6(1) & 0.50 & 0.96 \\
40 & 0.1626(2) & 0.48(5)  & 0.75 & 0.87 \\
60 & 0.16279(7)& 0.41(2)  & 1.12 & 0.25 \\
70 & 0.16288(6)& 0.39(2)  & 1.30 & 0.05 \\
 \end{tabular}
\caption{\label{ta2} Fitting results of the expression (\protect\ref{fit2})
to the scaled data shown in Fig.\protect\ref{fig2}. $\chi^2$ is per degree
of freedom and $Q$ is the quality of the fit.}
\end{table}

We have also fitted the expression
\begin{equation}
f(z)=a_1 (a_2-z), \label{fit2}
\end{equation}
to our scaled data (the constant $a_3$ has been absorbed into $a_2$), i.e.
neglecting the $\ln \ln$ term in Eq.(\ref{scale}).
The results are given in Table \ref{ta2}. We find values for
$a_1$ that differ somewhat from the predicted value of $1/(2\pi)$. 
Note that as far as the quality of the fit $Q$ (cf. Table \ref{ta2}) is 
concerned, Eq.(\ref{fit2}) is as good a description of the scaled
data in the range $-56 \leq z \leq -35$ as Eq.(\ref{fit}). However,
the fit value $a_1 \approx 0.1626$ does not agree as well with the predicted
value \cite{BREZIN} $1/(2\pi)$.
The inclusion of 
the $\ln \ln $ correction term yields 
much better agreement between the value of $a_1$ and
the universal value $1/(2\pi)$ 
predicted by the low temperature expansion (\ref{scale}). 

Let us conclude with the discussion of the possibility of a 
Kosterlitz--Thouless
transition in the two--dimensional O(3) model. Such a transition has 
been suggested by Seiler and Patrasciou \cite{SEILER} and its existence
has been subject to various Monte Carlo investigations\cite{SOKAL,ALLES}.
Since we observe finite--size scaling for the helicity modulus
we conclude from our scaled data (cf. Fig.\ref{fig2})
that the helicity modulus in
the limit of an infinite lattice
vanishes at all temperatures. Namely, by fixing the temperature and increasing
the lattice size we move along the scaling curve in the direction of 
increasing $z$ until the helicity modulus vanishes. The existence of a
Kosterlitz-Thouless
transition, however, would imply
a finite value for the helicity modulus below a
critical temperature even for an infinite lattice\cite{NELSON}. 
Our confirmation of the validity of the low temperature expansion together
with the strong indication of our scaled data that the helicity modulus 
on an infinite lattice vanishes, make a
Kosterlitz--Thouless transition in the two--dimensional O(3) model extremely
unlikely.

In conclusion we have computed numerically the finite--size scaling function
of the helicity modulus of the two--dimensional O(3) model and have found
very good agreement with the results of the low temperature expansion. We have
estimated the ranges of validity of the low temperature expansion
in the corresponding variables. Our results suggest that there is no 
Kosterlitz--Thouless transition in this model.

N.S. would like to thank S. Burnett for valuable discussions.


\begin{references}
\bibitem{POLYAKOV} A. M. Polyakov, Phys. Lett. {\bf 59B}, 79 (1975).
\bibitem{LOWT1} E. Br\'ezin and J. Zinn--Justin, Phys. Rev. Lett. {\bf 36}
691 (1976).
\bibitem{LOWT} E. Br\'ezin and J. Zinn--Justin, Phys. Rev.{\bf B14}, 3110 
(1976).
\bibitem{MERMIN} N. D. Mermin and H. Wagner, Phys. Rev. Lett. {\bf 17}, 1133
(1966).
\bibitem{THOULESS} J. M. Kosterlitz and D. J. Thouless, J. Phys. {\bf C6}, 1181
(1973), J. Phys. {\bf C7}, 1046 (1974).
\bibitem{SEILER} A. Patrascioiu and E. Seiler, Nucl. Phys. 
{\bf B (Proc. Suppl.) 30}, 184 (1993); A. Patrascioiu and E. Seiler, 
hep--lat/9508014.
\bibitem{SOKAL} S. Caracciolo, R. G. Edwards, A. Pelisseto, and A. D. Sokal,
Phys. Rev. Lett. {\bf 75}, 1891 (1995).
\bibitem{ALLES} B. All\'es, A. Buonanno, and G. Cella, hep--lat/9608002
\bibitem{NELSON} D. R. Nelson and J. M. Kosterlitz, Phys. rev. Lett. {\bf 39},
1201 (1977); J. V. Jose, L. P. Kadanoff, S. Kirkpatrick, and D. R. Nelson,
Phys. Rev. {\bf B16}, 1217 (1977).
\bibitem{BREZIN} E. Br\'ezin, E. Korutcheva, Th. Jolicoeur, and 
J. Zinn--Justin, J. Stat. Phys. {\bf 70}, 583 (1993).
\bibitem{WOLFFCL} U. Wolff, Phys. Rev. Lett. {\bf 62}, 361 (1989).
\bibitem{CHAKRAVARTY} S. Chakravarty, Phys. Rev. Lett. {\bf 66}, 481 (1990),
J. Rudnick and D. Jasnow, Phys. Rev. {\bf B16}, 2032 (1977).
\bibitem{HASENFRATZ} P. Hasenfratz and F. Niedermayer, Nucl. Phys. {\bf B337}
233 (1990).
\bibitem{WOLFF} U. Wolff, Nucl. Phys. {\bf B334} 581 (1990).
\bibitem{APOSTOLAKIS} J. Apostolakis, C. F. Baillie, and G. C. Fox, Phys.
Rev. {\bf D43} 2687 (1991).
\bibitem{RECIPE} W. H. Press, B. P. Flannery, S. A. Teukolsky, and 
W. T. Vetterling, {\it Numerical Recipes} (Cambridge University Press,
Cambridge, 1988).
\bibitem{NEUHAUS} T. Neuhaus, Nucl. Phys. {\bf B (Proc. Suppl.) 34}, 667 
(1994); hep--lat/9608043.
\bibitem{CAFFAREL}M. Caffarel, P. Azaria, B. Delamotte, and D. Mouhanna,
Europhys. Lett. {\bf 26}, 493 (1994).
\bibitem{HASEN}P. Hasenfratz, M. Maggiore, and F. Niedermayer, Phys. Lett.
{\bf B245}, 522 (1990); P. Hasenfratz and F. Niedermayer, 
Phys. Lett. {\bf B245}, 529 (1990).
\bibitem{BILLOIRE} A. Billoire, Phys. Rev. {\bf B54}, 990 (1996).
\end{references}
\end{document}